\documentclass[twoside]{dis04p}
 
\def\runauthor{}
\def\shorttitle{}
\def\ds{D^{\ast}}
\def\d0{D^{0}}
\def\dspm{{\ds}^{\pm}}

\def\GeV{\,\textrm{GeV}~}

\def\MeV{\,\textrm{MeV}}
\def\g2{\,\textrm{$GeV^2$}~}

\def\ptds{$p_{\perp}^{\ds}$}
\def\pt{p_{\!\perp}}
\def\ptrang{$p_{\perp}^{\ds}>1.35$\,GeV}
\def\etarang{$|\eta^{\ds}|<1.6$}
\def\stat{\,\textrm{(stat.)}} 
\def\syst{\,\textrm{(syst.)}}
 
\newcommand{\dstarplus}{D^{*+}}
\newcommand{\dstprplus}{D^{*'+}}
 
\begin{document}
 
\title{Charm hadron spectroscopy         
with ZEUS
}

%\author{URI KARSHON on behalf of the ZEUS Collaboration}
%\author{U.~Karshon \footnote{Supported by the Israel Science Foundation and the
%U.S.-Israel Binational Science Foundation} \\
%                       For the ZEUS Collaboration  
%\footnote{On behalf of the ZEUS Collaboration }  
%      }
\maketitle
 
\vspace*{-0.2cm}
\begin{center}
%\address{U.~Karshon \footnote{Supported by the Israel Science Foundation and the
       Uri Karshon \footnote{Supported by the Israel Science Foundation and the
U.S.-Israel Binational Science Foundation} \\
 
         Weizmann Institute of Science, Rehovot, Israel \\
                       For the ZEUS Collaboration \\
%E-mail: uri.karshon@weizmann.ac.il }
\end{center}
 
\vspace*{+0.5cm}
 
\abstracts{                                                           
%The production of excited charm mesons has been observed with the
%ZEUS detector at HERA.
                       Neutral orbitally excited P-wave charm mesons
have been reconstructed in the $D^{*\pm}\pi^{\mp}$ final state and
the charm-strange meson $D^{\pm}_{s1}(2536)$ was found in the
$D^{*\pm}K^0_s$ final state.                                      
A search for radially excited charm mesons in the $D^{*\pm}\pi^+\pi^-$
final state has also been performed.                                     
%The strange pentaquark candidate $\theta^+ (1530)$ has been seen at high photon virtuality,
%$Q^2$,              in the $K_S^0 p$ and
%$K_S^0\bar p$ spectra with a width consistent with experimental resolution.
A search for a charm pentaquark state near 3.1\GeV was made in the decay mode
 $D^{*\pm}p^{\mp}$. Using more than 40,000 reconstructed $D^*$ mesons, no resonance
structure was observed.                                                                        
}
 
\vspace*{-0.8cm}
\section{Introduction}                    
 
\vspace*{-0.1cm}
The years 2003-2004 brought new life to hadron spectroscopy. New unexpected               
narrow states were found in various places: in the $D_s$ sector, a higher charmonium
state $X(3872)$ and, of most interest, new pentaquark candidates have been claimed by
various experiments. The most established one by now is the exotic baryon state
$\theta^+ (1530)$ decaying into $K^+ n$ or $K^0 p$ with strangeness=+1, as predicted by
Diakonov et al.~\cite{Diakonov} at the top of a $SU(3)$ anti-decuplet of baryons.
The minimal quark composition of this new state is $u u d d \bar s$.
 
In March 2004, the H1 Collaboration at HERA reported~\cite{H1}
                                the    observation of a narrow state in the 
 $D^{*\pm}p^{\mp}$ spectrum at 3.1\GeV and attributed it to the charm pentaquark      
$\theta^0_c (u u d d \bar c)$. In this talk preliminary ZEUS results are presented on 
charm spectroscopy of states decaying into a $D^{*\pm}$ plus other hadrons.

 \vspace*{-0.1cm}
\section{Charm tagging for spectroscopy}
 
 \vspace*{-0.1cm}
The charmed meson $D^{*\pm}$ has been reconstructed                                     
                                                    via its decay chain                                          
      $D^{*+}~\rightarrow~D^0~\pi_S^+~\rightarrow~( K^- \pi^+ )~\pi_S^+$~(+c.c.).  
Fig.~1(a) shows the      mass difference distribution, $\Delta M~=~M(K\pi\pi_S)-M(K\pi)$,
%                    in the  $M(K\pi)$ mass region of the $D^0$
                            in the kinematic range
\ptrang~and \etarang , where $p_{\perp}$ is the transverse momentum
%          and  $\eta =-\ln\tan(\theta /2)$ is the pseudorapidity.   
           and  $\eta$                      is the pseudorapidity.   
%The polar angle, $\theta$, is defined with respect to the proton beam direction.
The region $1.83 < M(K\pi) < 1.90\GeV$ was used for low \ptds~and a somewhat wider
region was applied for high \ptds.
 The plot includes all the ZEUS data collected during            
  1995-2000 and corresponds to an integrated luminosity of 126.5 pb$^{-1}$.
A clear $\dspm$ signal is seen. The combinatorial background is estimated
from the wrong charge combinations, where both $D^0$ tracks have the
same charge and $\pi_S$ has the opposite charge.                                  
For the following charm spectroscopy studies,
$\dspm$ candidates were defined as events with $0.144 < \Delta M < 0.147\GeV$.
In this range (shaded band in Fig.~1(a)) a signal of $42730\pm 350~\dspm$                         
            mesons was found after wrong charge
            background subtraction. This corresponds to a statistical precision
of better than $1\%$.
In Fig.~1(b) only deep inelastic scattering (DIS) events are considered with a scattered
%electron energy $> 8$~GeV and $Q^2~>~1\GeV^2$. The signal is cleaner
electron energy $> 8\GeV$ and $Q^2 > 1\g2$. The signal is cleaner
but $\approx 4.5$ times smaller than in the inclusive case, with 
$N(D^*) = 9697\pm 145$.
 
 \vspace*{-0.2cm}
 
\begin{figure}[!thb]
%\vspace*{-3.0cm}
\begin{center}
%\special{psfile=prel_fig1.eps voffset=-60 vscale=40
%hscale= 40 hoffset=10 angle=0}
%\special{psfile=figs/kpi_5q.q2_1.eps voffset=-60 vscale=40
%hscale= 40 hoffset=10 angle=0}
%\centerline{\epsfxsize=2.9in\epsfbox{kim_mephi_lep.ps}}
%\hspace*{-0.5cm}\psfig{figure=prel_fig1.eps,height=15.0cm}
%\hspace*{-1.0cm}\psfig{figure=kpi_5q.q2_1.eps,height=15.0cm}
%\centerline{\epsfxsize=2.9in\epsfbox{prel_fig1.eps}}
%\hspace*{-0.5cm}
%\vspace*{+3.0cm}
\resizebox{14.0pc}{!}{\includegraphics{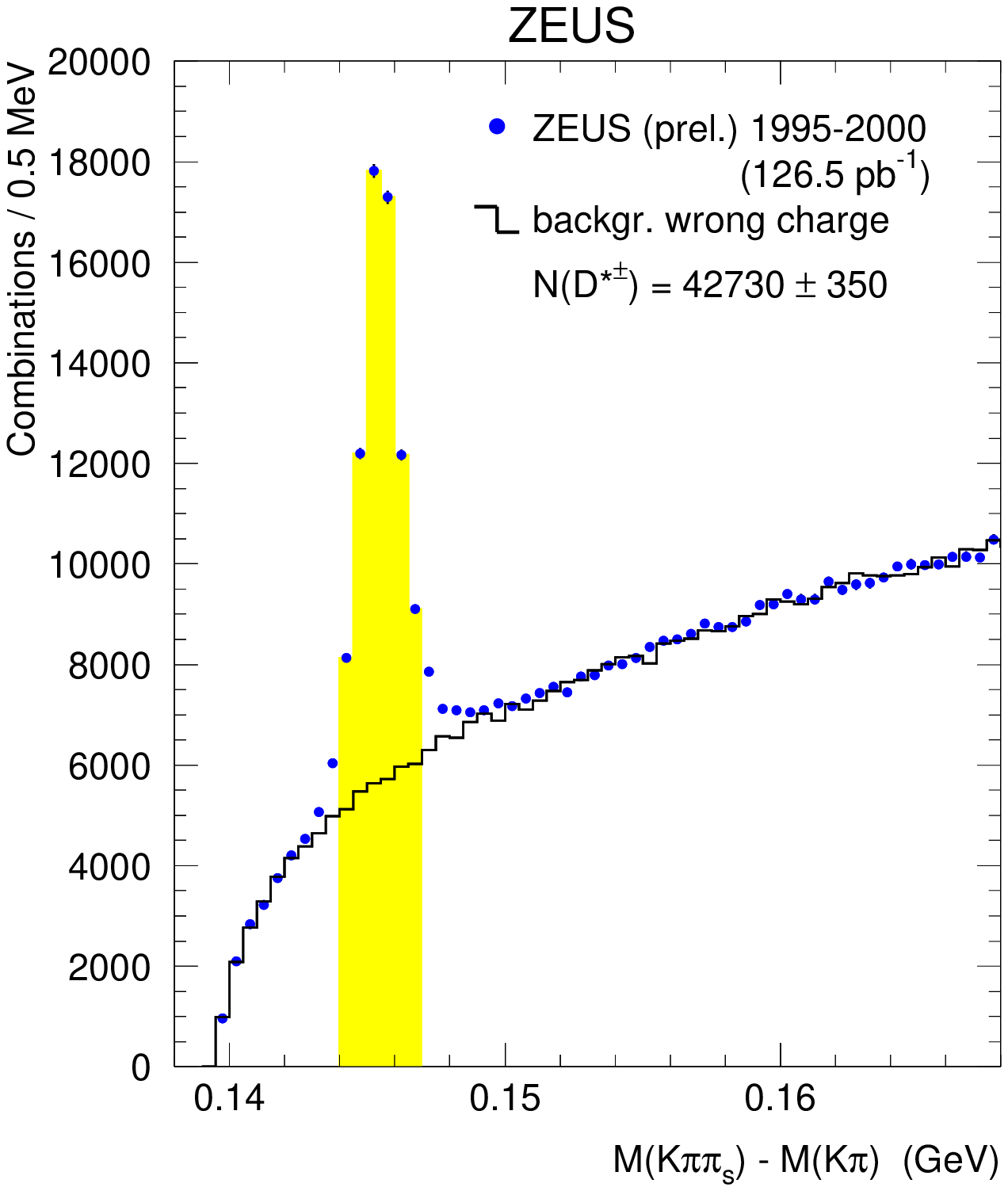}}
%\vspace*{-12.0cm}\hspace*{+0.9cm}\resizebox{14pc}{!}{\includegraphics{kpi_5q.q2_1.eps}}
\vspace*{-12.0cm}\hspace*{+0.4cm}\resizebox{14pc}{!}{\includegraphics{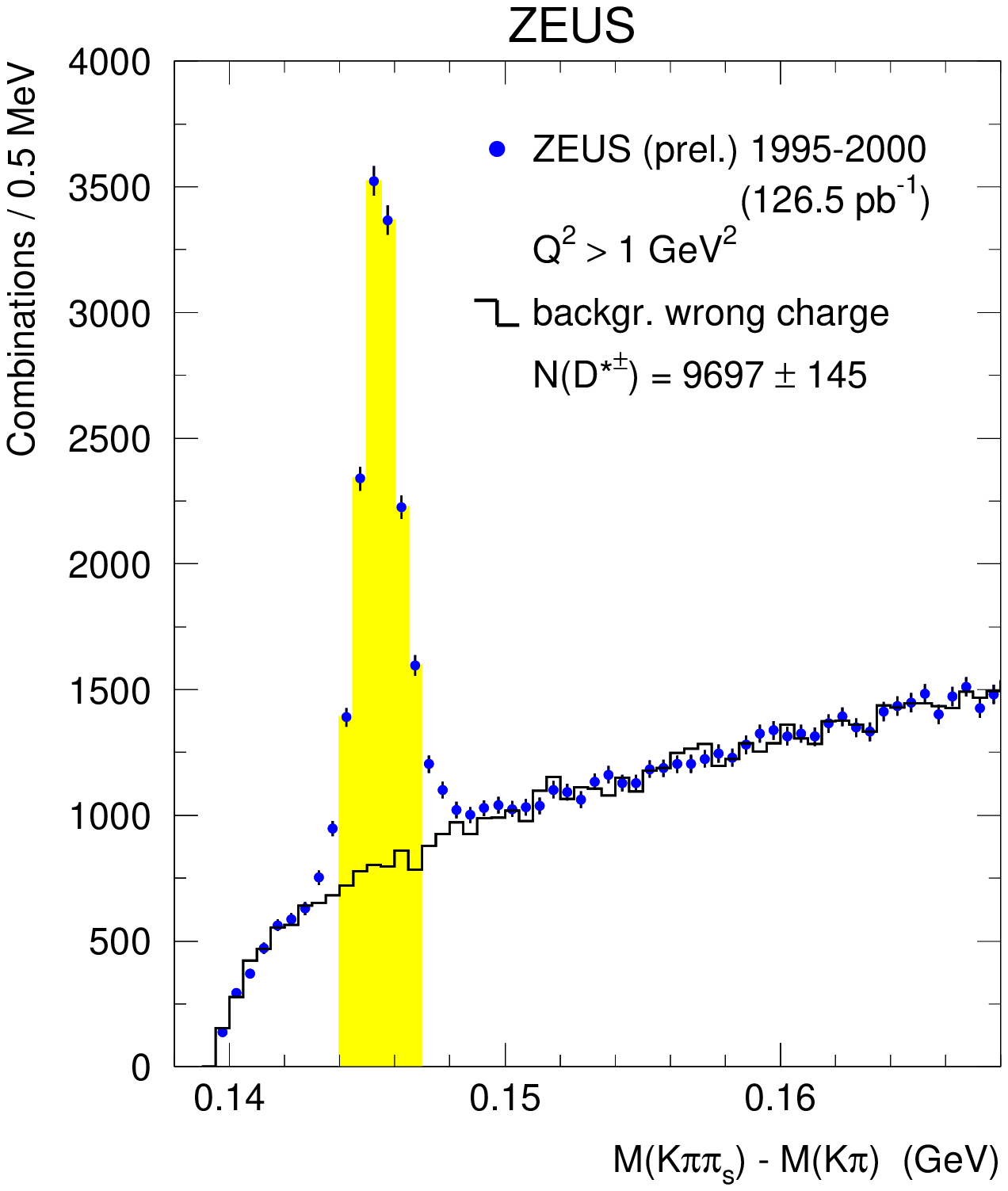}}
%\vspace*{+8.8cm}
 \vspace*{+11.8cm}
  \caption{(a)
  $M(K\pi\pi_S)-M(K\pi)$ distribution in the $D^0$ mass region (dots).
  The        histogram is the distribution for wrong charge combinations.
%  (b) Same as (a) for DIS events $Q^2 > 1$\GeV$^2$.
   (b) Same as (a) for DIS events $Q^2 > 1\g2$.
   }
\end{center}
\end{figure}
 
\vspace*{- 7.6cm}\hspace*{ 0.8cm}{\tiny  (a)}
\vspace*{  7.6cm}\hspace*{-0.8cm}
 
\vspace*{- 8.0cm}\hspace*{ 7.3cm}{\tiny   (b)}
\vspace*{+ 8.0cm}\hspace*{-7.3cm}
  
\vspace*{-1.7cm}
 
\section{Excited charm mesons}
 
\vspace*{-0.1cm}
 
$D^0_1(2420)$ and $D^{*0}_2(2460)$ mesons were                   
               reconstructed~\cite{pwave} via their decays 
to $D^{*\pm}\pi^{\mp}_4$, followed by the $D^{*\pm}$ decays,
   $D^{*+}\to D^0\pi^+_S\to (K^-\pi^+ )\pi^+_S(+c.c.)$.             
                        Fig.~2(a) shows the ``extended" mass difference distribution,
$M(K\pi\pi_S\pi_4)-M(K\pi\pi_S)+M(\ds)$, where $M(\ds)$ is the PDG     $\dspm$
mass~\cite{PDG}. A clear excess is seen around the                   
 $D_1^0$ and $D^{*0}_2$ mass region.               
No enhancement is seen for wrong charge combinations, where the
$D^*$ and $\pi_4$ have the same charges.
The solid curves in Figs.~2(a-b) are an unbinned likelihood fit to two
Breit-Wigner shapes with masses and widths fixed to the nominal 
$D^0_1$ and $D^{*0}_2$ values~\cite{PDG}, convoluted with a Gaussian function and
multiplied by helicity spectrum functions for $J^P=1^+$ and $2^+$ states, respectively.
                              The background shape was parametrised by the form
 $x^\alpha \cdot exp(-\beta \cdot x + \gamma \cdot x^2)$, where   
  $x =                                          
 M(K\pi\pi_S\pi_4)-M(K\pi\pi_S)-M(\pi)$. The fitted curves describe the distribution 
reasonably well, except for a narrow enhancement near
$2.4\GeV$ (Fig.~2(b)).
In Fig.~2(c), a similar fit is shown with an additional Gaussian-shaped resonance
with free mass and width.                                                             
                                                  The fit yielded $211\pm 49$ entries
for the narrow enhancement with mass value $2398.1\pm 2.1\stat ^{+1.6}_{-0.8}\syst $
\MeV. The width was consistent with the          resolution expected from the tracking
detector.                   The enhancement may
indicate a new excited charm meson, a result of an interference effect or a statistical
fluctuation.
The number of reconstructed 
$D^0_1$ and $D^{*0}_2$ mesons in the 3-resonance fit are $526\pm 65$ and $203\pm 60$,
respectively.

%\vspace*{-1.0cm}
 
$D_{s1}^{\pm}(2536)$ mesons were reconstructed~\cite{ds1}
 via the             $\dspm K^0_S$ decay mode                
                                   with $K^0_S~\to~\pi^+~\pi^-$.
$K^0_S$ candidates were identified by using pairs of oppositely charged tracks with
$\pt > 0.2\GeV$.
                                  A      clean                                         
                                               $K^0_S\to \pi_3\pi_4$                  
                          signal was      extracted after
applying        $V^0$-finding cuts~\cite{ds1}.  $K^0_S$ candidates with
$0.480 < M(\pi_3\pi_4) < 0.515\GeV$ were kept for the $D_{s1}^{\pm}$  reconstruction.
%For each                                             
%$D_{s1}^{\pm}$ candidate,                           
%                       the extended mass difference,
                                          Fig.~2(d)   shows the          
effective $M(\dspm K^0_S)$ distribution in terms of
%$\Delta M^{ext}+M(D^{*+})_{PDG}+M(K^0)_{PDG}$, where
$\Delta M^{\rm ext}+M(D^{*+})_{\rm PDG}+M(K^0)_{\rm PDG}$, where
$\Delta M^{\rm ext}=
 M(K\pi\pi_S\pi_3\pi_4)-M(K\pi\pi_S)-M(\pi_3\pi_4)$  and                     
$M(D^{*+})_{\rm PDG}$ ($M(K^0)_{\rm PDG}$) is the nominal
                       $\dspm$ ($K^0$) mass~\cite{PDG}.
A clear signal is seen at the $M(D^{\pm}_{s1})$ value.
                        The       curve is an unbinned likelihood
fit to a Gaussian resonance plus background of the form $A(\Delta M^{\rm ext})^B$. The fit
yielded $62.3\pm~9.3~D^{\pm}_{s1}$ mesons with                            
$M(D^{\pm}_{s1})=2534.2\pm 0.6\pm 0.5\MeV$, in rough agreement with the PDG value~\cite{PDG}.    
The last error is due to the uncertainty in    
$M(D^{*+})_{\rm PDG}$.                                                                       
The angular distribution of the 
$D_{s1}$ signal was studied via the                                                       
 helicity angle, $\alpha$,                      between the $K^0_S$ and $\pi_S$
momenta in the $\dspm$ rest frame. The $dN/d\cos\alpha$ 
            distribution                  was fitted to
($1+R\cos^2\alpha$). An  unbinned likelihood fit yielded 
$R = -0.53 \pm 0.32 \stat^{+0.05}_{-0.14}\syst $, consistent with the CLEO
value~\cite{CLEO}
 $R = -0.23^{+0.40}_{-0.32}$. Both measurements are consistent with $R=0$, i.e.               
$J^P=1^+$ for the 
$D_{s1}$ meson. However, the result presented here is also  consistent with $R=-1$, 
        i.e. $J^P=1^-$ or $2^+$~\cite{godfrey}.
 
 \vspace*{-1.0cm}
 
\begin{figure}[!thb]
%\begin{figure}[t]
\begin{center}    
 \hspace*{-0.5cm}
\vspace*{+3.0cm}
  \resizebox{14pc}{!}{\includegraphics{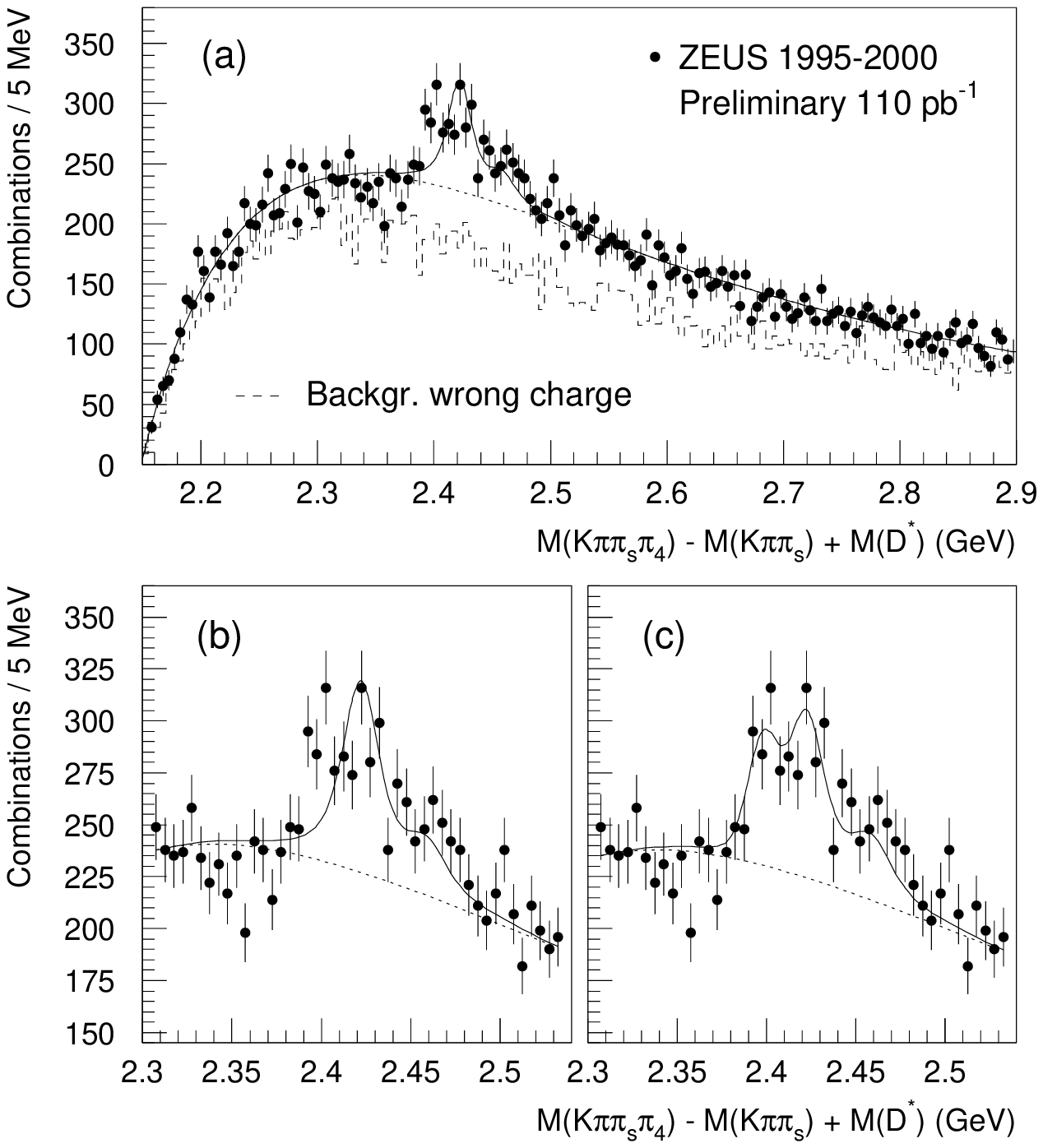}}      
\vspace*{-12.0cm}\hspace*{+0.9cm}
  \resizebox{14pc}{!}{\includegraphics{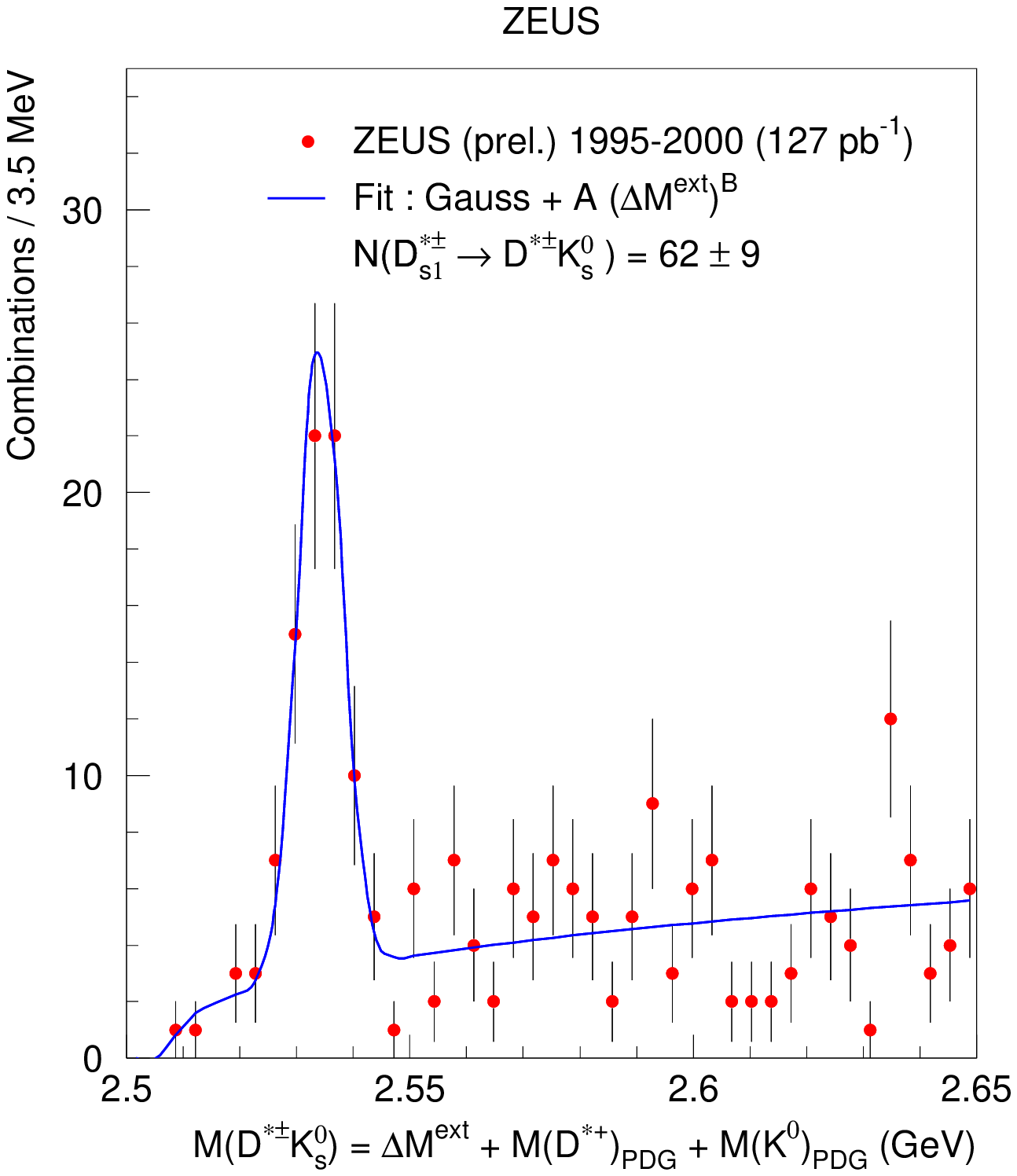}}      
 
 \vspace*{+ 8.8cm}
 
\caption{(a-c)~Extended mass difference distribution,                              
$M(K\pi\pi_S\pi_4)-M(K\pi\pi_S)+M(\ds)$ (dots).
The   histogram is for wrong charge combinations. The curves are from the unbinned   
likelihood fits. In (a) and (b) the solid curves are a fit to background parametrisation
and two Breit-Wigner distributions convoluted with a Gaussian function. In (c)
an additional Gaussian-shaped resonance near $2.4\GeV$ is assumed in the fit. The dotted
curves are fitted shapes of the combinatorial background.
(d)      Effective                                                                 
          $M(\dspm K^0_S)$ distribution (dots). The solid line is a                  
fit to a Gaussian resonance plus background of the form $A(\Delta M^{\rm ext})^B$.        
         }
\end{center}
\end{figure}
 
\vspace*{- 9.3cm}\hspace*{ 7.2cm}{\tiny   (d)}
\vspace*{+ 9.3cm}\hspace*{-7.2cm}
  
 \vspace*{-0.7cm}
 
Radially excited charm mesons,                          
$D^{*'}$,                               
with mass around      $2.6\GeV$ are predicted~\cite{radial}
to decay   to  $D\pi\pi$ or $D^*\pi\pi$.                                  
A narrow ($\Gamma < 15$~MeV) signal of $66\pm 14$ events            
was reported in    
 $M(D^{*\pm}\pi^+\pi^-)$             by DELPHI~\cite{DELPHI} at 
$2637$~MeV and interpreted as a   radially excited 
$D^{*'\pm}$. No evidence for this
state has been found by OPAL and CLEO~\cite{OPAL,CLEO2}.
$D^{*'\pm}$  candidates were               
                 reconstructed~\cite{pwave} from their decays to
         $D^{*\pm}\pi^+_4\pi^-_5$.    
No narrow resonance is seen in
                                  the extended mass difference
$M(K\pi\pi_S\pi_4\pi_5)-M(K\pi\pi_S)+M(\ds)$.                     
                  An upper limit for
%$R_{\dstprplus \to \dstarplus \pi^+\pi^-/\dstarplus}<2.3\%~~(95\%~~C.L.)$ is obtained
   the fraction of $\dspm$        originating from    
$D^{*'\pm}$ decays                                            
in the measured kinematic region is obtained
                            within a signal window    $2.59 <
M(D^{*'\pm}) < 2.67\GeV$, which covers theoretical predictions~\cite{radial} and the
DELPHI measurement~\cite{DELPHI}.
Extrapolating by a MC simulation to the full kinematic phase space
              and using the known $f(c\to D^{*+})$ value~\cite{LG},
a $D^{*'\pm}$ production limit of                                                             
 $f(c \to \dstprplus) \cdot B_{\dstprplus \to \dstarplus \pi^+ \pi^-}
                                          < 0.7 \%~(95\%~C.L.)$ is obtained.
A similar limit of  $0.9\%$ has been reported    by OPAL~\cite{OPAL}.

 \vspace*{-0.6cm}
\section{Search for a charm pentaquark}
 
 \vspace*{-0.1cm}
 
Various QCD models speculate that the existence of the strange
pentaquark $\theta^+ =~u~u~d~d~\bar s$ implies that heavy pentaquarks, such as
$\theta_c^0 = u u d d \bar c$, should also exist. Some models~\cite{Jaffe} predict
$M(\theta_c^0)\approx 2700$~MeV, which is too light to decay strongly to D mesons. Other
models~\cite{Karliner} predict 
$M(\theta_c^0) = 2985\pm 50$~MeV with $\Gamma(\theta_c)\approx 21$~MeV with a dominant
decay mode to $D^- p$ or $D^0 n$ (+c.c.).
If $M(\theta_c^0) > M(\dspm)+M(p) = 2948$~MeV, $\theta_c^0$ can decay to $\dspm p$.     
The H1 Collaboration reported recently~\cite{H1} evidence for a narrow resonance in the
 $D^{*\pm}p^{\mp}$ mass spectrum around 3.1\GeV and attributed it to the charm pentaquark.     
They find that $\approx 1\%$ of the $\dspm$ mesons originate from this state.
 
A search for narrow charm pentaquark                                                      
                        states was done in the $M(D^{*-}p)$(+c.c.) decay channel with the
full 1995-2000 data set.                 
Here charm pentaquark  candidates were formed by combining       $\dspm$ candidates
from Fig.~1 with a fourth track, assumed to be a proton, with a charge opposite to 
        the $\dspm$.
Proton and anti-proton tracks with momentum $P$ lie within a wide $(dE/dx)_p$ band: 
$0.3/P^2 + 0.8 < dE/dx < 1/P^2 + 1.2$.                                 
To suppress the large $\pi$/K background, two
proton selections were used:                       $P < 1.35$\GeV and
$(dE/dx)_p > max(1.3,0.3/P^2+0.8)$ or 
 $P > 2$\GeV and
$(dE/dx)_p < 1/P^2+1.2$.   
For each charm pentaquark candidate the   extended  mass difference,
$\Delta M^{\rm ext} = M(K\pi\pi_s p) - M(K\pi\pi_s)$, was calculated.
 
%\vspace*{-0.6cm}
Fig.~3 shows the $M(D^* p)=\Delta M^{\rm ext} + M(D^{*+})_{\rm PDG}$
distribution for the low- and high-$P$ proton selections.       
%where $M(D^{*+})_{PDG}$ is the nominal $\dspm$ mass~\cite{PDG}.
The upper plots include all candidates, while the lower plots have DIS events only.
The histograms are      
 $M(D^* p)$ distributions for like-sign combinations of $D^*$ and proton.
No narrow resonance is seen. 
In order to check if a charm pentaquark signal was not lost due to the  selection requirements
or     hidden in the combinatorial background, 
the cuts were varied.         The main
systematic checks were:           
varying the
$dE/dx$ requirements in both low- and high-$P$ proton selection; 
remove reflections from 
$D^0_1 ,  D^{*0}_2\to \dspm\pi^{\mp}$ decays; make all cuts as close as possible to
the H1 selection~\cite{H1}. No signal was seen in any of the selection variations.
 
To compare the measurement qualitatively with the H1 charm pentaquark 3.1\GeV 
                                                         signal~\cite{H1}, a naive
estimation of the expected signals was performed, assuming a rate of $1\%$ for
$\dspm$ mesons originating from the reconstructed charm pentaquark. Assuming that the
         signal  contributes            
30\% and 40\%, respectively,
to the low- and high-$P$ selections, 128 and 171 events are expected in these
mass distributions.
In Fig.~4 
                                           the       curves are minimal $\chi^2$
fits to the form $A(\Delta M^{\rm ext}-~m_p)^B exp[-(\Delta M^{\rm ext}-m_p)C]$, where $m_p$
is the proton mass. The fake Gaussian signals with the above estimated number of events
and the mass and width of the H1 signal are shown on top of the fitted curves.
                                                          The data constrain the
uncorrected fraction of $\dspm$ mesons originating from a hypothetic              
$\theta^0_c\to D^{*\pm} p^{\mp}$ resonance at 3.1\GeV to be
well below $1\%$.

 \vspace*{-0.5cm}
 
\begin{figure}[!thb]
%\begin{figure}[t]
\begin{center}    
%\vspace*{+3.0cm}
\hspace*{-0.5cm}
\resizebox{10pc}{!}{\includegraphics{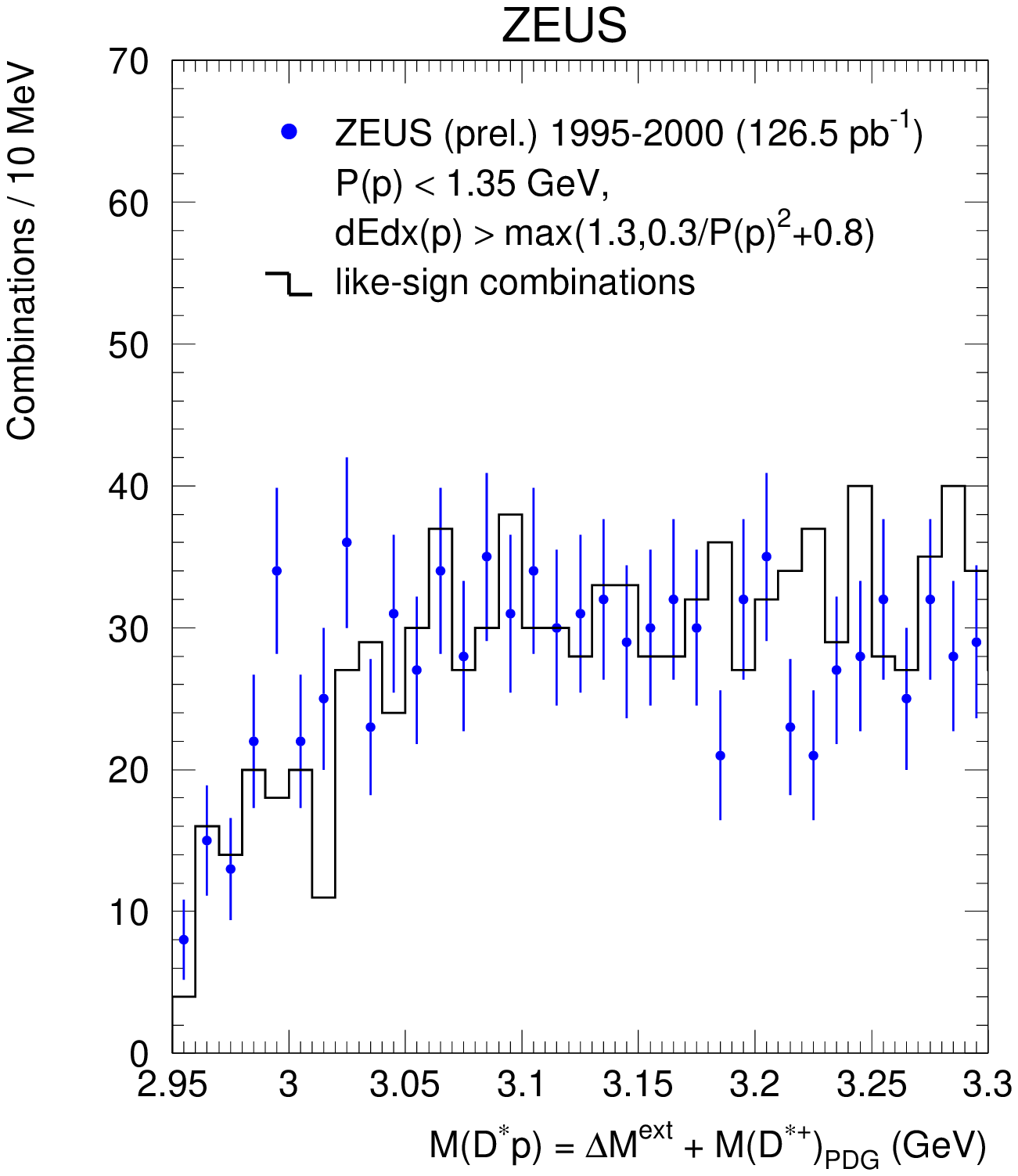}}      
%\vspace*{-12.0cm}
\hspace*{+0.9cm}
\resizebox{10pc}{!}{\includegraphics{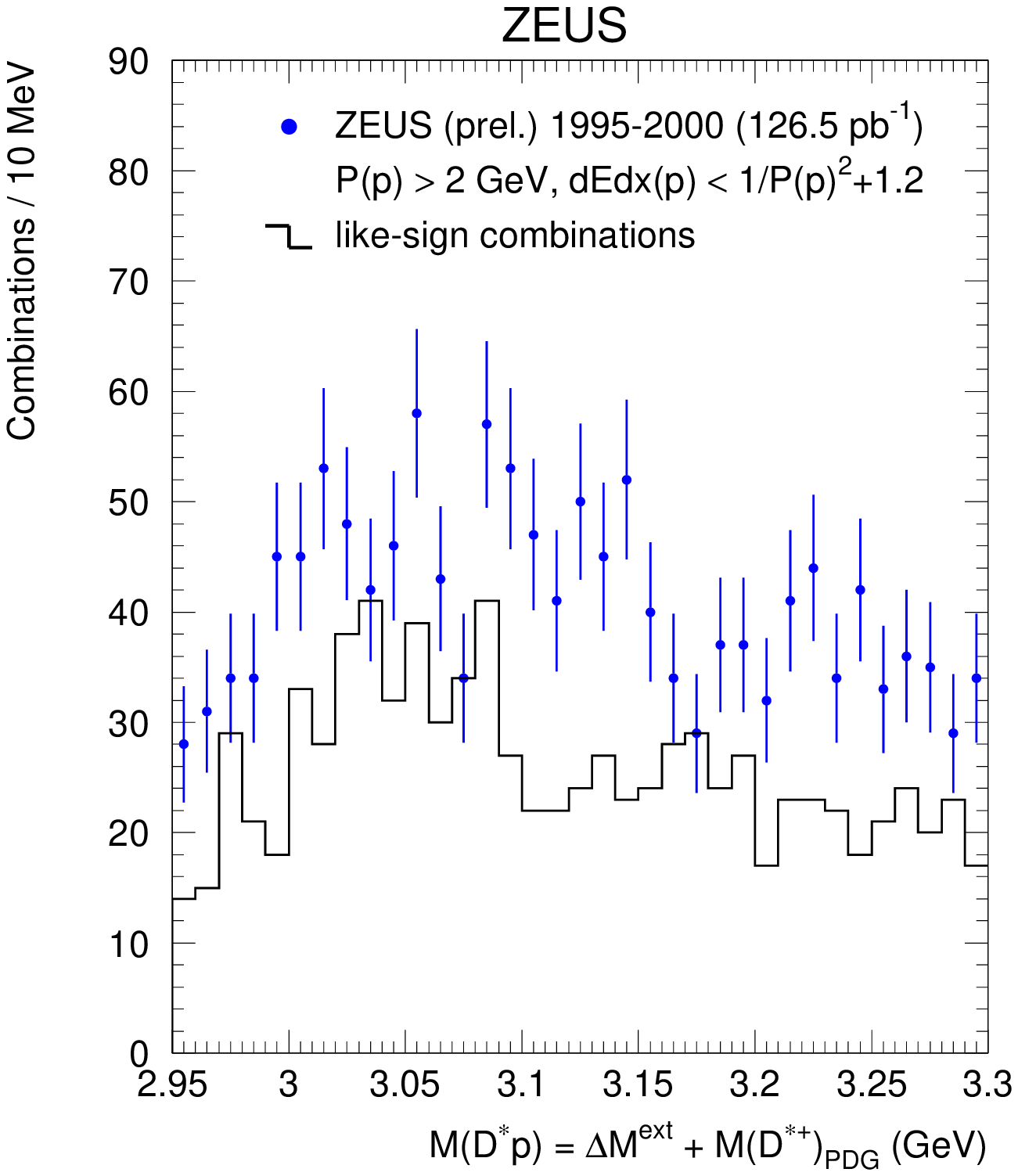}}      
%\vspace*{+9.0cm}
 
\hspace*{-0.5cm}
 
\resizebox{10pc}{!}{\includegraphics{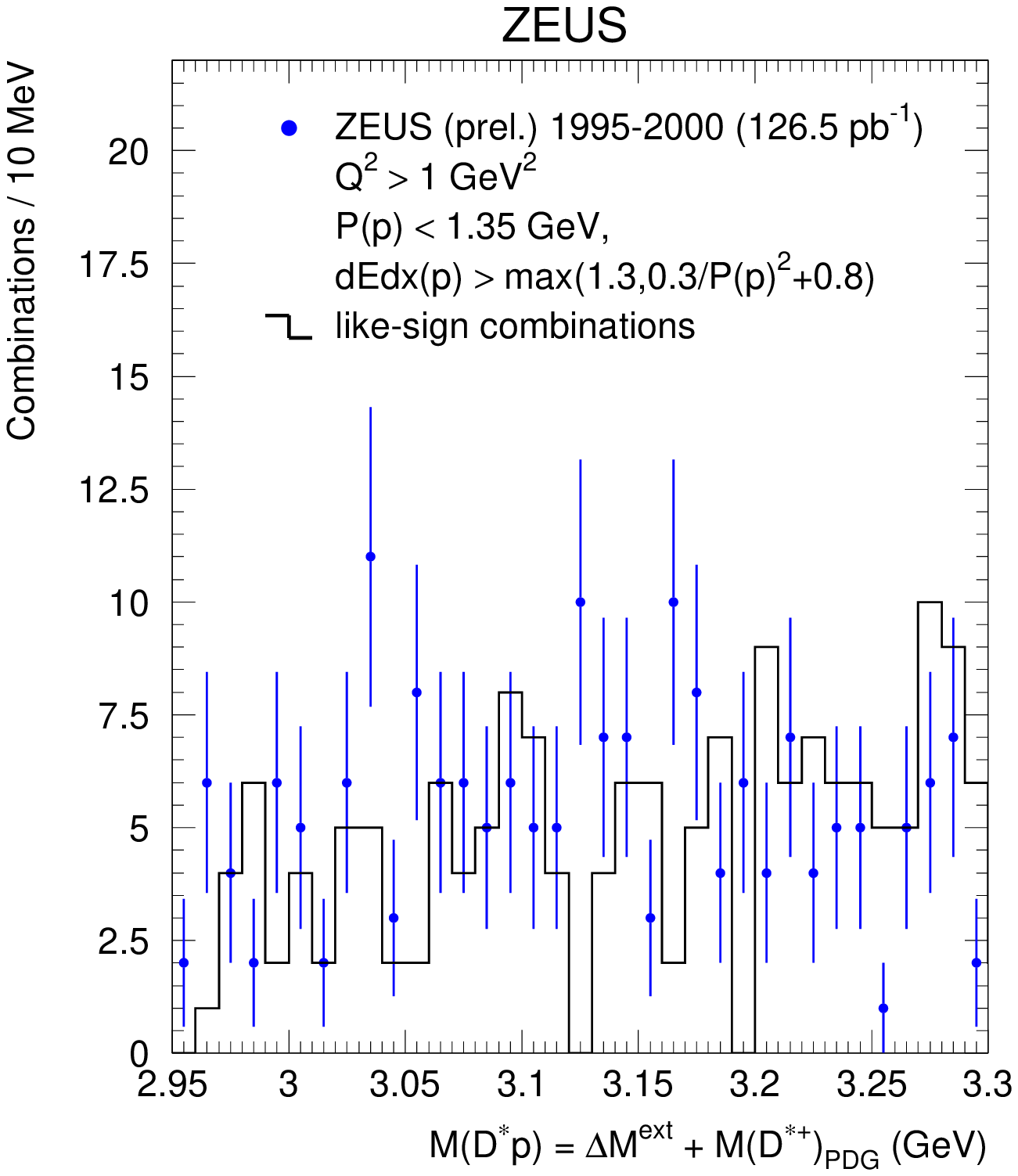}}
%\vspace*{-12.0cm}
\hspace*{+0.9cm}
\resizebox{10pc}{!}{\includegraphics{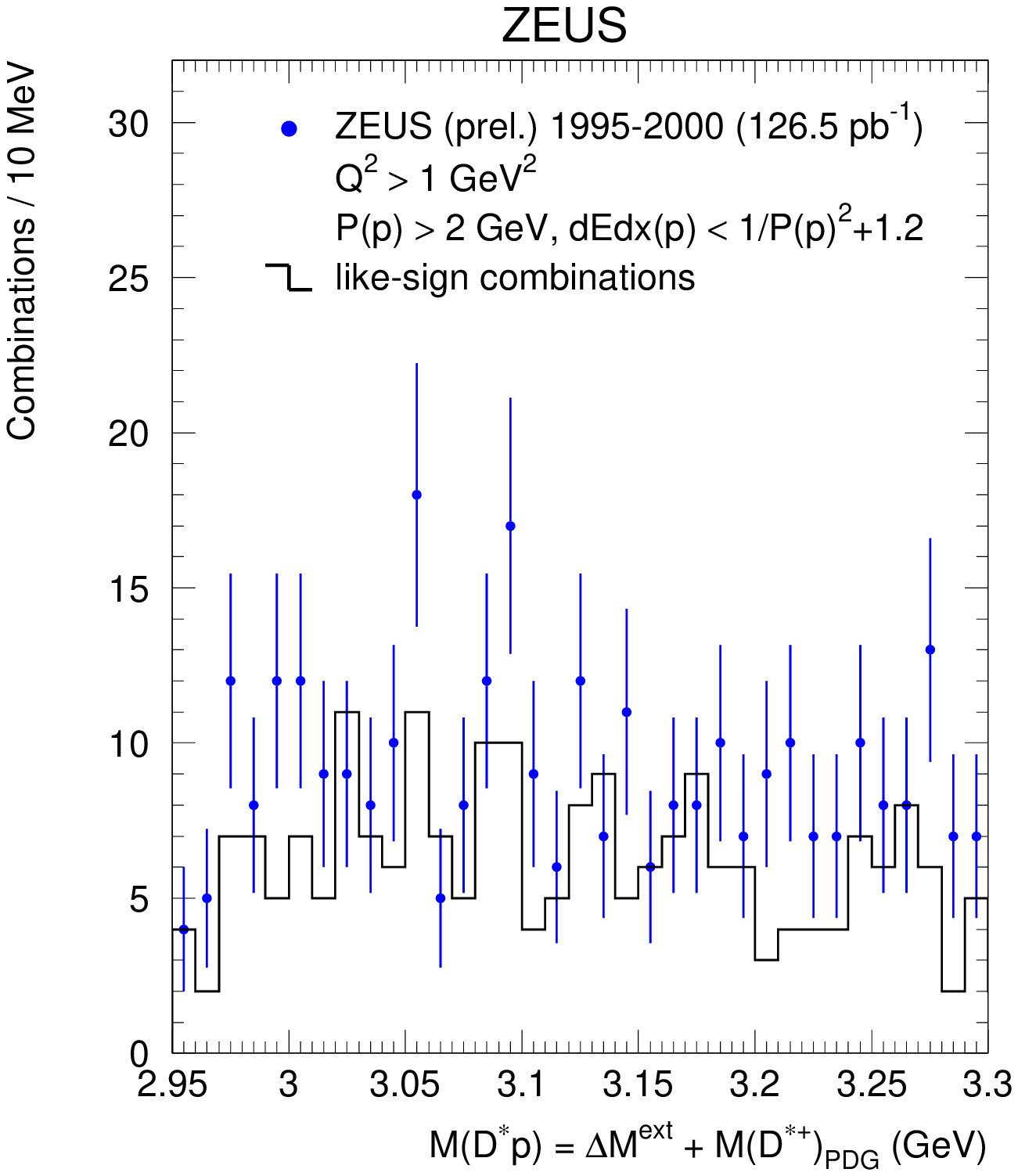}}
 
 \vspace*{-0.3cm}
\caption{$M(D^* p)$ distributions for charm pentaquark candidates (dots) with low-$P$ (up-left) and
 high-$P$ (up-right) proton selections. The histograms are             
    $M(D^* p)$ distributions for like-sign $D^*$ and p combinations.             
The lower plots are the same for DIS events.
         }
\end{center}
\end{figure}

 \vspace*{+0.2cm}
 
\section{Conclusions}
 
 \vspace*{-0.1cm}
The ZEUS 1995-2000 data sample was used to study charm spectroscopy of states
decaying into a $\dspm$ plus other hadrons. The P-wave charm mesons 
$D^0_1(2420)$, $D^{*0}_2(2460)$ and                           
$D^{\pm}_{s1}(2536)$ are clearly seen. No evidence is found for the radially
excited state
$D^{*'\pm}(2637)\to                              
  D^{*\pm}\pi^+\pi^-$ seen by DELPHI.                 
No resonance structure is seen in the                             
$M(D^{*\pm} p^{\mp})$ spectra.
The data is not compatible with a contribution from $\theta_c$ to the overall $D^*$ rate
of $\approx 1\%$, as reported by the H1 Collaboration.
 
 \vspace*{-0.5cm}
 
\begin{figure}[!thb]
%\begin{figure}[t]
\begin{center}    
%\vspace*{+3.0cm}
\hspace*{-0.5cm}
\resizebox{10pc}{!}{\includegraphics{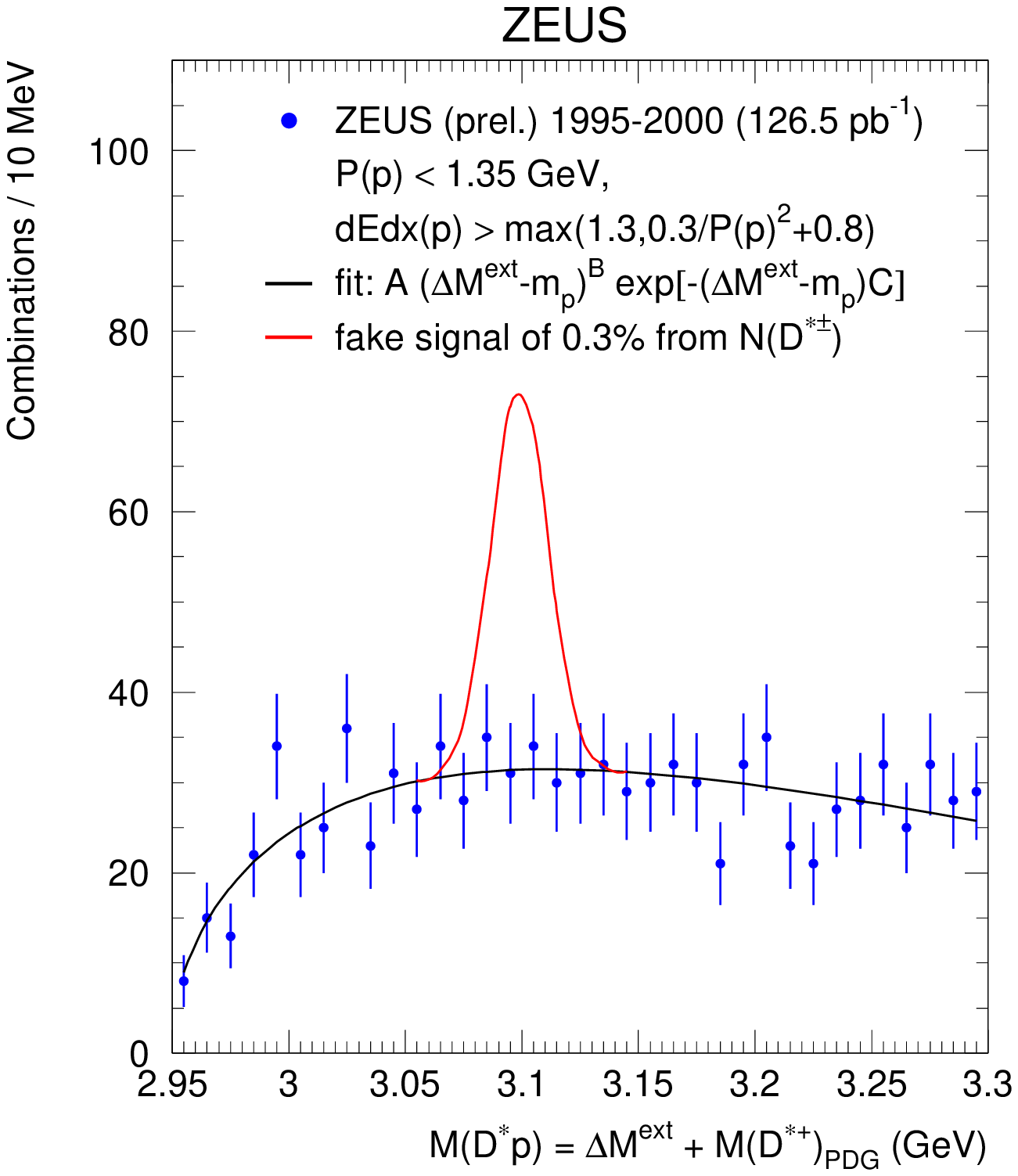}}      
%\vspace*{-12.0cm}
\hspace*{+0.9cm}
\resizebox{10pc}{!}{\includegraphics{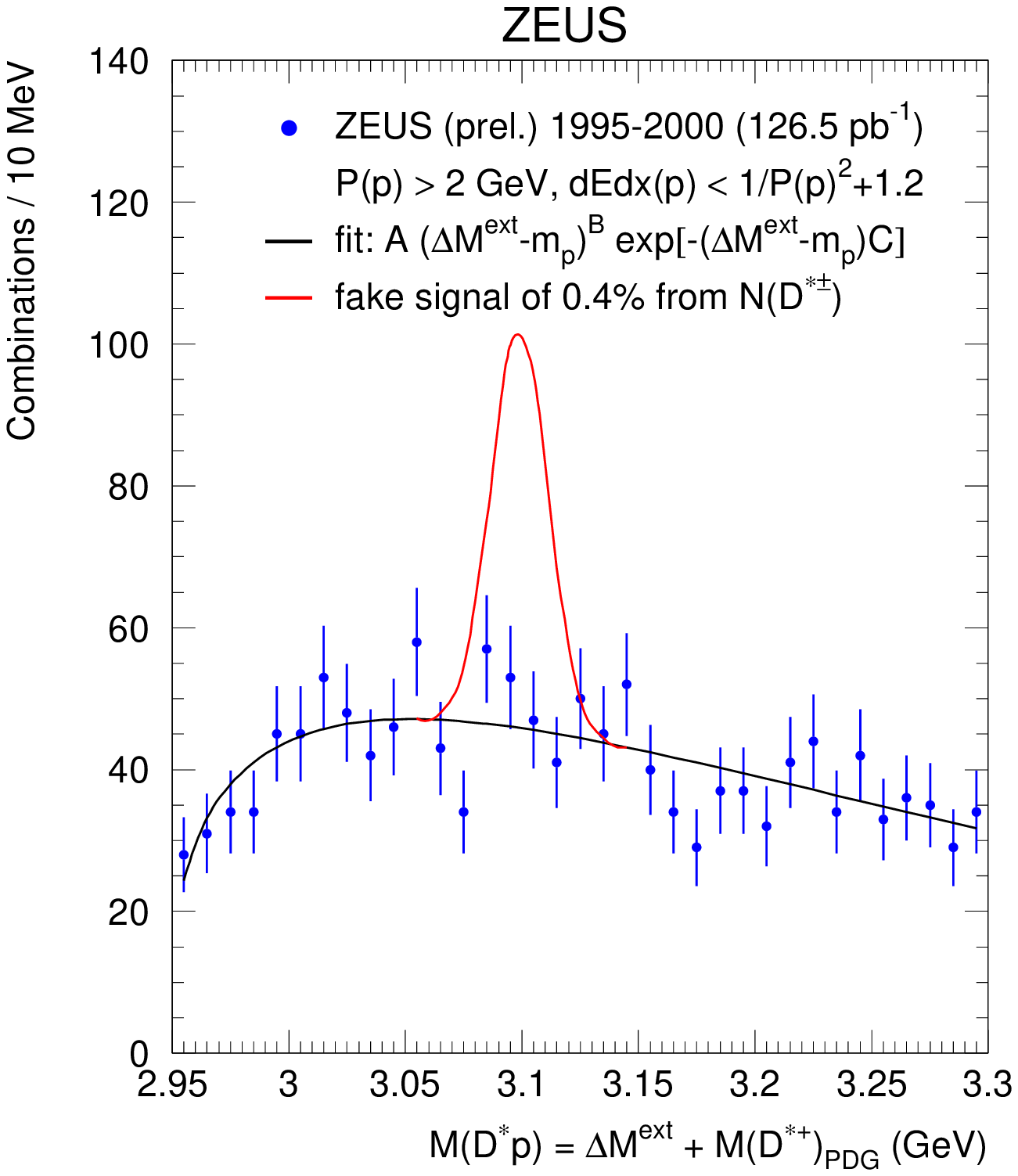}}      
%\vspace*{+9.0cm}
 
\hspace*{-0.5cm}
\resizebox{10pc}{!}{\includegraphics{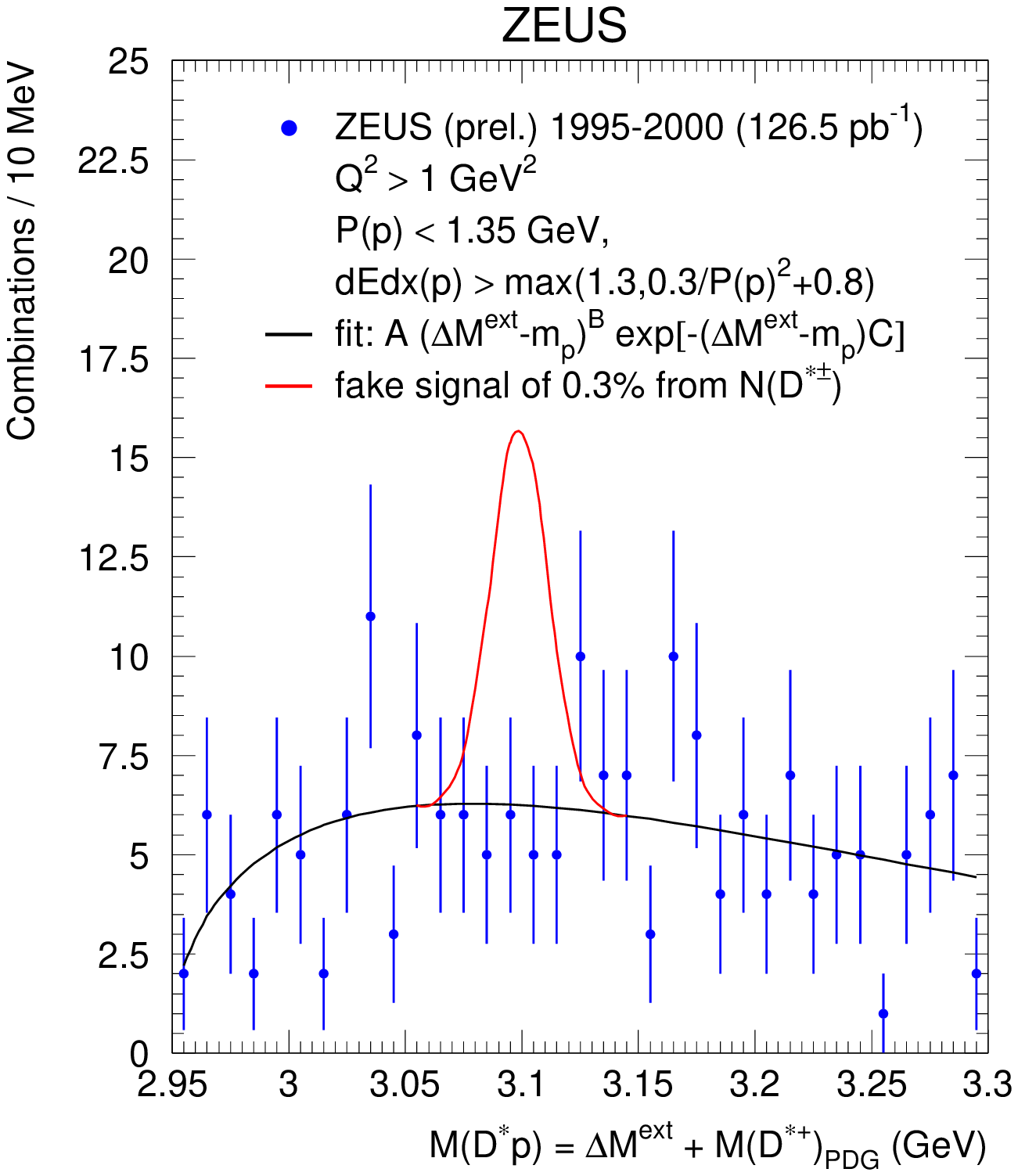}}
%\vspace*{-12.0cm}
\hspace*{+0.9cm}
\resizebox{10pc}{!}{\includegraphics{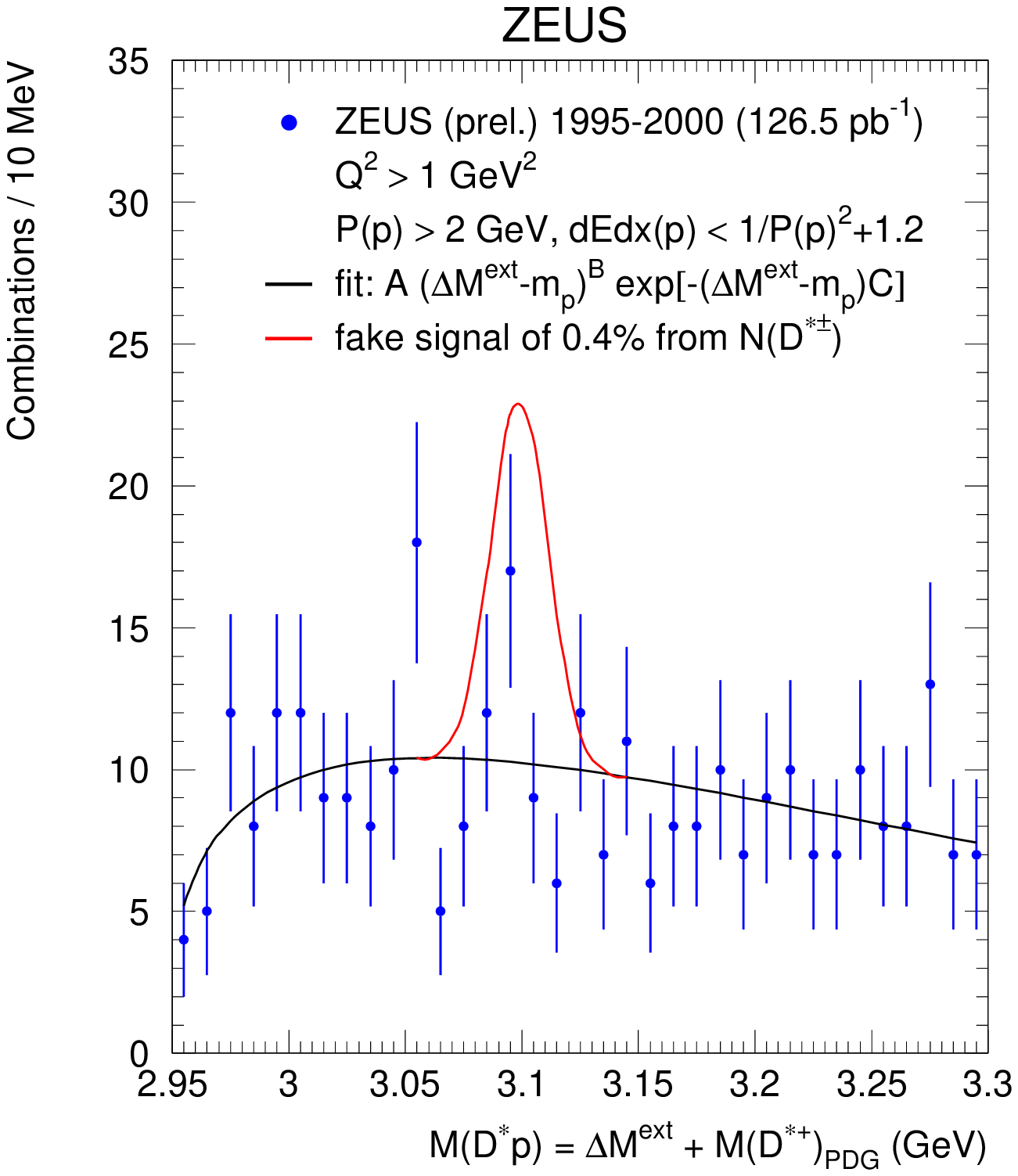}}
 
%\vspace*{+12.0cm}
\caption{$M(D^* p)$ distributions for charm pentaquark candidates (dots) with low-$P$ (up-left) and
 high-$P$   (up-right) proton selections. The curves are described in the text.
The lower plots are          for DIS events.
         }
\end{center}
\end{figure}

\vspace*{+0.3cm}


\begin{thebibliography}{0}
 
\bibitem{Diakonov} D.~Diakonov, V.~Petrov and M.~V.~Polyakov,
Z. Phys. A {\bf359} (1997) 305.   
 
\bibitem{H1} H1 Collaboration, A.~Aktas et al.,        
Phys. Lett. B {\bf588} (2004) 17.   
 
\bibitem{pwave} ZEUS Collaboration, XXX Int. Conf. on High Energy Physics,
ICHEP2000, Osaka, Japan, July-August 2000, paper 448.
 
\bibitem{PDG} Particle Data Group, K.~Hagiwara et al., 
Phys. Rev. D {\bf66} (2002) 10001.   
 
%\bibitem{PDG98} Particle Data Group, C.~Caso et al., 
%Eur. Phys. J {\bf C3} (1998) 1.   
 
\bibitem{ds1} ZEUS Collaboration, Int. Europhysics Conf. on High Energy Physics,
Budapest, Hungary, July 2001, abstract 497.
 
\bibitem{CLEO} CLEO Collaboration, J.~P.~Alexander et al., 
Phys. Lett. B {\bf303} (1993) 377.  
 
\bibitem{godfrey} S.~Godfrey and R.~Kokoski,           
Phys. Rev. D {\bf43} (1991) 1130.   
 
\bibitem{radial} D.~Ebert et al.,                          
Phys. Rev. D {\bf57} (1998) 5663.   
 
\bibitem{DELPHI} DELPHI Collaboration, P.~Abreu et al.,   
Phys. Lett. B {\bf426} (1998) 231.  
 
\bibitem{OPAL} OPAL Collaboration, G.~Abbiendi et al.,                      
 Eur. Phys. J {\bf C20} (2001) 445. 
 
\bibitem{CLEO2}                                                                       
J.~L.~Rodriguez (CLEO Collaboration) hep-ex/9901008 and Proceedings of ``Heavy
Quarks at Fixed Targets", Fermilab, October 1998, ed. H.~W.~K.~Cheung and J.~N.~Butler
(1999).
 
\bibitem{LG} L.~Gladilin, Preprint hep-ex/9912064, 1999.
 
\bibitem{Jaffe} R.~L.~Jaffe and F.~Wilczek,  
Phys. Rev. Lett. {\bf91} (2003) 232003.  
 
\bibitem{Karliner} M.~Karliner and H.~Lipkin, hep-ph/0307343 (2003).
 
 
\end{thebibliography}
\end{document}